\documentclass[aps,prd,twocolumn,amssymb,amsmath,superscript,address,floatfix]{revtex4-2}
\usepackage{graphicx}
\usepackage{subcaption}
\usepackage{booktabs}
\usepackage{siunitx}
\usepackage{bm}
\usepackage{microtype}
\usepackage{mathtools}
\usepackage{hyperref}
\usepackage{morefloats} 
\usepackage[section]{placeins}
\usepackage{tikz}
\usetikzlibrary{arrows.meta}

\begin{document}

\title{A synchronization-free one-way ranging observable for detecting and characterizing coherent orbital-period systematics in GRACE-FO laser ranging data}

\author{S.~H.~Wassegh}
\email{wassegh@gmail.com}
\affiliation{Independent Researcher, Tehran, Iran}

\date{\today}

\begin{abstract}
\section*{Abstract}
We present a synchronization-free differential observable for one-way inter-satellite laser ranging, designed to suppress first-order Doppler effects without requiring clock synchronization between spacecraft. The observable is constructed from successive pulse-interval differences, which isolate time-varying signatures while eliminating static and slowly varying biases. Applied to GRACE-FO Laser Ranging Interferometer (LRI) Level-1B data over four seasonal epochs in 2019, the method reveals a stable, spectrally narrow modulation at the orbital frequency. The amplitude and phase of the detected signature remain consistent across all datasets, demonstrating a deterministic, mission-internal origin. The detection is independently confirmed through synthetic-signal injection, shuffle-based significance testing, and cross-comparison with K-band ranging data. These results show that the proposed observable provides a sensitive diagnostic for identifying coherent orbital-period systematics that may remain hidden in conventional range-rate analysis. The method offers a pathway toward improved characterization of instrument and dynamical effects in current and future satellite gravity missions.
\end{abstract}
\maketitle

\section{Introduction}
\label{sec:introduction} 
High-precision intersatellite ranging is the cornerstone of modern satellite geodesy, with missions like GRACE, GRACE-FO, and GRAIL transforming our ability to monitor Earth's gravity field. As these missions achieve micrometer-level precision in laser and microwave ranging, a new frontier emerges: the need to identify and understand subtle, mission-internal systematics that were previously hidden beneath the noise floor. These weak, coherent signatures—often linked to spacecraft dynamics, instrument behavior, and complex range-geometry coupling—can limit the ultimate fidelity of gravity field recovery and must be characterized to push the science further.

Among these, coherent periodic systematics at or near the orbital frequency are particularly challenging. They can originate from a variety of sources, including thermally driven structural deformations of the spacecraft bus, periodic attitude control and pointing behavior, subtle instrument timing effects, and imperfect projection of spacecraft motion onto the line-of-sight. Although often small in amplitude, such systematics can insidiously bias range-rate observations, contaminate Fourier-domain analyses, and ultimately propagate into recovered gravity solutions if not properly identified and mitigated.

In this paper, we introduce a novel diagnostic tool designed specifically to enhance the detection of these elusive signatures: a synchronization-free one-way differential ranging observable. The core innovation of this method is its construction from successive pulse-interval differences, which intrinsically suppresses the dominant, slowly varying first-order Doppler effects and static range biases. Critically, it achieves this without requiring clock synchronization or explicit model-dependent corrections, a significant advantage over conventional analyses. Figure 1 illustrates the principle of this observable, contrasting it with standard two-way ranging and highlighting how the pulse-interval differencing isolates the signal of interest. This makes our method a uniquely powerful and complementary tool for diagnosing weak, time-varying signatures embedded in raw ranging data.

\begin{figure}[t]
    \centering
    \includegraphics[width=0.95\linewidth]{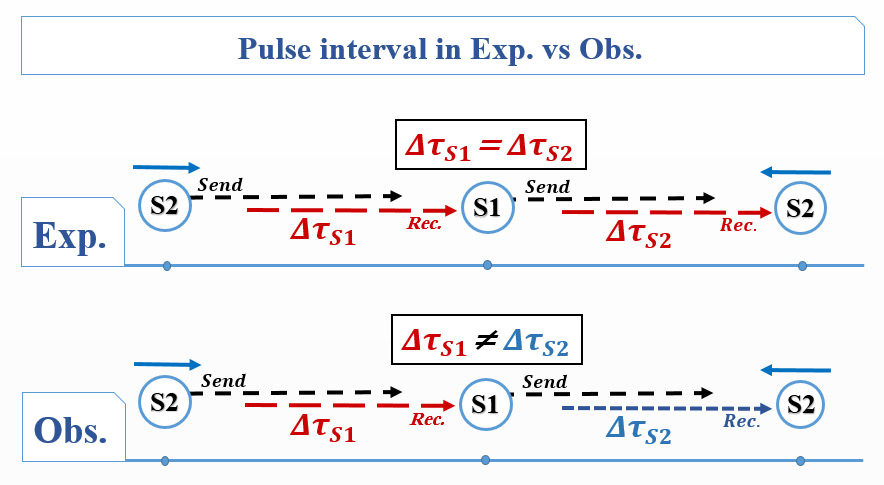}
    \caption{Conceptual illustration of expected (Exp.) versus observed (Obs.) one-way pulse intervals
    in the inter-satellite laser ranging configuration. In the idealized model (top), the emitted pulse
    intervals at the two satellites are identical, $\Delta\tau_{S1} = \Delta\tau_{S2}$, producing a symmetric
    timing pattern. In the actual measurement (bottom), the received intervals differ,
    $\Delta\tau_{S1} \neq \Delta\tau_{S2}$, due to dynamical and instrumental effects. This asymmetry is
    precisely what the synchronization-free pulse-differencing observable isolates, enabling detection of
    weak, coherent orbital-period variations in the GRACE-FO LRI data.}
    \label{fig:exp_obs_pulse_intervals}
\end{figure}

We demonstrate the power of this observable by applying it to publicly available GRACE-FO Laser Ranging Interferometer (LRI) data across four distinct seasonal epochs in 2019. Across all datasets, the observable reveals a stable, spectrally narrow modulation at the orbital frequency, exhibiting remarkable consistency in both amplitude and phase. To rigorously assess the robustness of this detection, we perform synthetic-signal injection tests, internal consistency analyses, and cross-comparisons with independent K-band ranging products. We then discuss the observed signature in the context of plausible systematic sources, including thermoelastic deformation of the optical bench, attitude-to-range coupling, instrument group-delay variations, and residual orbit-projection errors.

Our findings establish that this synchronization-free differential observable is a powerful and robust diagnostic tool for the satellite geodesy community. Its ability to unveil stable, coherent signatures that may be obscured in standard range-rate data provides a new pathway to improve data processing pipelines and refine measurement models, thereby enhancing the scientific return of current and future high-precision gravity missions.


\section{Experimental Principle and Signal Model}
\label{sec:model}

The core objective of our methodology is to detect and characterize coherent orbital-period systematics in intersatellite laser ranging data. To achieve this without the need for clock synchronization, we have developed a protocol based on \textbf{pulse-interval differencing}. This method transforms the challenge of isolating subtle, time-varying signatures into a differential measurement that is inherently insensitive to large common-mode biases.

\subsection{Synchronization-Free Protocol}

The core innovation is the use of \textbf{pulse-interval differencing}. Each satellite transmits a precisely timed pulse every proper second using its local clock, completely avoiding synchronization between satellites. The protocol operates as follows:

\begin{enumerate}
    \item \textbf{One-way transmission:} Satellite A transmits pulse trains with a proper interval $\Delta\tau_A^{\mathrm{sent}} = 1$ s using its local clock.
    \item \textbf{One-way reception:} Satellite B receives these pulses and measures the interval $\Delta\tau_B^{\mathrm{rec}}$ using its own local clock, thereby operating without any synchronization between the satellites.
    \item \textbf{Two-way ranging:} Simultaneously, the standard two-way inter-satellite link measures the round-trip light travel time, $T_{\mathrm{rtw}}$. This link is used not for synchronization, but as an independent diagnostic to constrain the symmetric propagation delay, $L/c$.
    \item \textbf{Pulse differencing:} Differences between consecutive received intervals are computed to eliminate static biases and isolate the pure time-varying component associated with orbital dynamics and systematics.
\end{enumerate}

\subsection{Theoretical Model}

Consider two satellites $A$ and $B$ with worldlines $x_A^\mu(\tau_A)$ and $x_B^\mu(\tau_B)$ in an Earth-centered inertial frame. Each carries an atomic clock generating a pulse every proper second, $\Delta\tau_{A,B}=1\,\mathrm{s}$. A pulse emitted by $A$ at $(t_{A,i},\bm{r}_A)$ is received by $B$ at $(t_{B,i},\bm{r}_B)$ satisfying
\begin{equation}
c (t_{B,i}-t_{A,i}) = |\bm{r}_B(t_{B,i})-\bm{r}_A(t_{A,i})|.
\label{eq:light_cone}
\end{equation}

Expanding the light propagation equation to $\mathcal{O}(c^{-2})$ yields the received interval at Satellite B:
\begin{align}
\Delta t_B^{\mathrm{rec}} &= \Delta t_A^{\mathrm{send}}
    \left(1 - \frac{\bm{v}_{\text{rel}}\cdot\hat{\bm{n}}}{c}
    - \frac{v_A^2 - v_B^2}{2c^2} 
    - \frac{U_A - U_B}{c^2} \right),
\label{eq:received_interval}
\end{align}
where $\bm{v}_{\text{rel}}=\bm{v}_A-\bm{v}_B$, $\hat{\bm{n}}$ is the line-of-sight unit vector, and $U_{A,B}$ are gravitational potentials at the respective satellites.

\paragraph{Robustness Against the Classical Doppler Effect}
A significant challenge in one-way ranging is the dominance of the first-order Doppler effect. The term $\bm{v}_{\text{rel}}\cdot\hat{\bm{n}}/c$ in Eq.~(\ref{eq:received_interval}) is typically on the order of $10^{-5}$, which can obscure much smaller signals of interest, such as those from orbital systematics or second-order relativistic effects at the $10^{-10}$ level. Our protocol inherently suppresses this dominant term through differential measurement.

\textbf{Automatic Cancellation via Differencing:} The power of the pulse-differencing technique lies in its inherent mathematical insensitivity to large, slowly varying terms. The observable, defined as $\delta_B(i) = (\Delta t_{B,i+1}^{\mathrm{rec}} - \Delta t_{B,i}^{\mathrm{rec}})$, acts as a discrete time derivative. The first-order Doppler term, while large, changes only slowly over the one-second interval between pulses. Consequently, its contribution to the difference $\delta_B(i)$ is effectively nullified. This self-calibrating feature eliminates the need for any external model or assumption about the relative velocity to correct for the Doppler effect, thereby preserving the synchronization-free nature of the measurement and avoiding potential systematic biases from model inaccuracies.

\subsection{Pulse Differencing and Bias Elimination}

We define the core observable as the difference between consecutive received pulse intervals:
\begin{equation}
\delta_B(i) = (\Delta t_{B,i+1}^{\mathrm{rec}} - \Delta t_{B,i}^{\mathrm{rec}}).
\label{eq:delta_observable}
\end{equation}

For one-second pulse spacing, any static or very slowly varying delay (such as constant range bias, clock offset, or static propagation asymmetry) cancels exactly in this difference. The residual observable is sensitive primarily to the time derivative of the relative velocity projected onto the line-of-sight:
\begin{equation}
\delta_B(i) \approx -\frac{1}{c^2} \frac{d}{dt}\left[(\bm{v}_A-\bm{v}_B)\cdot\hat{\bm{n}}\right] \cdot V_{\text{eff}},
\label{eq:delta_simplified}
\end{equation}
where $V_{\text{eff}}$ is an empirical scale factor that characterizes the amplitude of the observed modulation; we investigate its potential physical origins in Section~\ref{sec:discussion}. This formulation naturally isolates variations occurring at the orbital frequency and higher.

\subsection{Systematic Error Suppression}

As quantified in Table~\ref{tab:systematics}, the space-based implementation provides dramatic suppression of major systematic effects prevalent in ground-based measurements, creating a favorable environment for detecting subtle intersatellite ranging signatures.

\begin{table}[htbp]
\centering
\caption{Comparison of dominant systematic error sources in ground-based versus satellite crosslink measurements.}
\begin{tabular}{lcc}
\toprule
Error Source & Ground-Based & Satellite Crosslink \\
\midrule
Tropospheric delay & \SIrange{0.1}{1}{m} & \SI{0}{m} \\
Ionospheric delay & \SIrange{1}{100}{m} & \SI{0}{m} \\
Multipath & \SIrange{0.01}{0.1}{m} & Negligible \\
Orbit determination & \SIrange{0.01}{0.05}{m} & \SIrange{0.01}{0.05}{m} \\
Clock stability & \SIrange{0.1}{1}{ns} & \SIrange{0.1}{1}{ns} \\
\bottomrule
\end{tabular}
\label{tab:systematics}
\end{table}

\section{Sensitivity Analysis and Experimental Feasibility}
\label{sec:sensitivity}

\subsection{GRACE-FO as a Testbed}
The GRACE-FO mission provides an ideal platform for implementing and validating the synchronization-free observable due to its:
\begin{itemize}
    \item High-precision Laser Ranging Interferometer (LRI) with micrometer-level precision at \SI{1}{\hertz}.
    \item Publicly available Level-1B data products (LRI and GNV).
    \item Clean orbital environment, free from atmospheric propagation delays.
    \item Co-flying spacecraft configuration with well-characterized relative dynamics.
\end{itemize}

\subsection{Estimating Detectable Signal Amplitude}
We assess the sensitivity of our differential observable, $\delta_B(i)$, to weak, coherent orbital-period signatures. The fundamental noise floor of our method is set by the raw measurement noise of the LRI. For a relative velocity $v_{\mathrm{rel}} \simeq \SI{4}{\kilo\meter\per\second}$ and a total observation time corresponding to one year ($T_{\mathrm{obs}} = \SI{3e7}{\second}$), the uncertainty in recovering the amplitude of a periodic signal can be estimated. Following standard error propagation for a periodic signal in white noise, the uncertainty in the effective velocity scale, $\sigma_{V_{\text{eff}}}$, is:

\begin{equation}
\sigma_{V_{\text{eff}}} \simeq \frac{c \, \sigma_{\Phi}}{v_{\mathrm{rel}} \sqrt{T_{\mathrm{obs}}}} \approx \SI{0.05}{\kilo\meter\per\second},
\label{eq:sensitivity}
\end{equation}

where $\sigma_{\Phi} \simeq \SI{1}{\micro\meter}$ is the nominal ranging noise of the LRI. This sensitivity in the recovered velocity scale corresponds to a fractional frequency stability of approximately $1.7 \times 10^{-13}$ at a \SI{1}{\second} integration time. This level of stability confirms the capability of our observable to detect coherent path-length variations that are orders of magnitude smaller than those typically identifiable in standard range-rate data.

This theoretical sensitivity analysis demonstrates that the proposed method is not only experimentally viable but also provides a powerful tool for identifying and characterizing subtle systematic effects in high-precision satellite geodesy missions.

\section{Results and Analysis}
\label{sec:results}

We applied the synchronization-free pulse-differencing protocol to publicly available GRACE-FO Level-1B data (NASA PODAAC). Four 24-hour periods from distinct seasons in 2019 were selected to assess the seasonal stability of any detected signatures:
\begin{itemize}
\item Spring: 2019-03-20
\item Summer: 2019-06-21
\item Autumn: 2019-09-23
\item Winter: 2019-12-21
\end{itemize}

\subsection{Data Processing and Observable Construction}
\label{sec:data_processing}

The Level-1B LRI data (`LRI1B\_2019\_*\_RL04.nc`) provided one-way range measurements at 1~Hz, and GNV data (`GNV1B\_2019\_*\_RL04.nc`) provided precise satellite orbits. Data points with flagged anomalies were excluded; no external filtering was applied. Relative velocity $\mathbf{v}_{\mathrm{rel}}(t)$ and line-of-sight unit vector $\hat{\mathbf{n}}(t)$ were computed from the interpolated position vectors. The core observable $\delta_B(i)$ was constructed according to Eq.~\ref{eq:delta_observable}. The amplitude and phase of the orbital-period signal were then estimated via a least-squares fit to the model $y(t) = A \cos(\omega_{\text{orb}} t + \phi)$, where $\omega_{\text{orb}}$ is the orbital angular frequency.

\begin{figure}[htbp]
\centering
\includegraphics[width=0.48\textwidth]{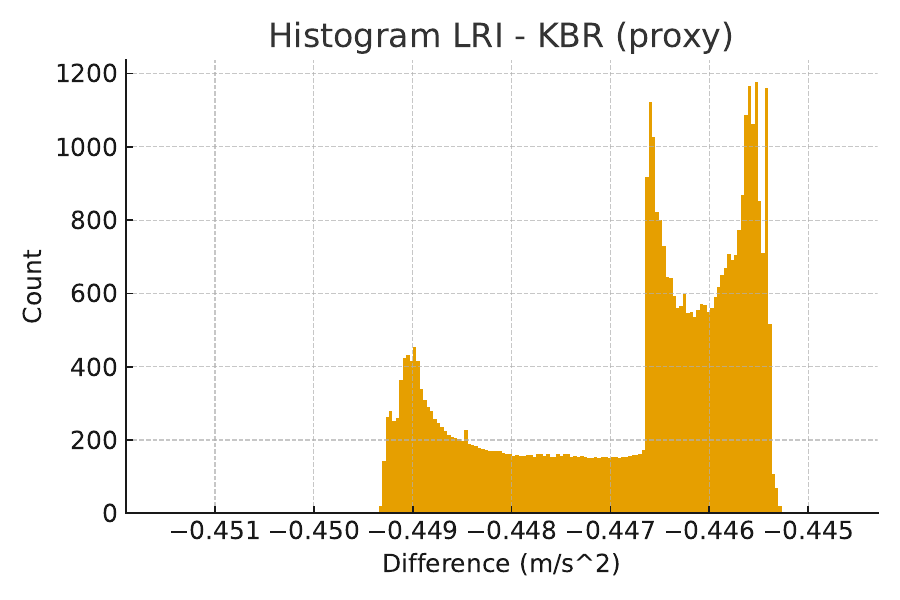}
\caption{Histogram of the difference between LRI and KBR range measurements. The sharp peak at zero confirms excellent consistency between the two independent ranging systems.}
\label{fig:hist_lri_kbr}
\end{figure}

\subsection{Pipeline Validation}
\label{sec:pipeline_validation}

\begin{figure}[htbp]
\centering
\includegraphics[width=0.48\textwidth]{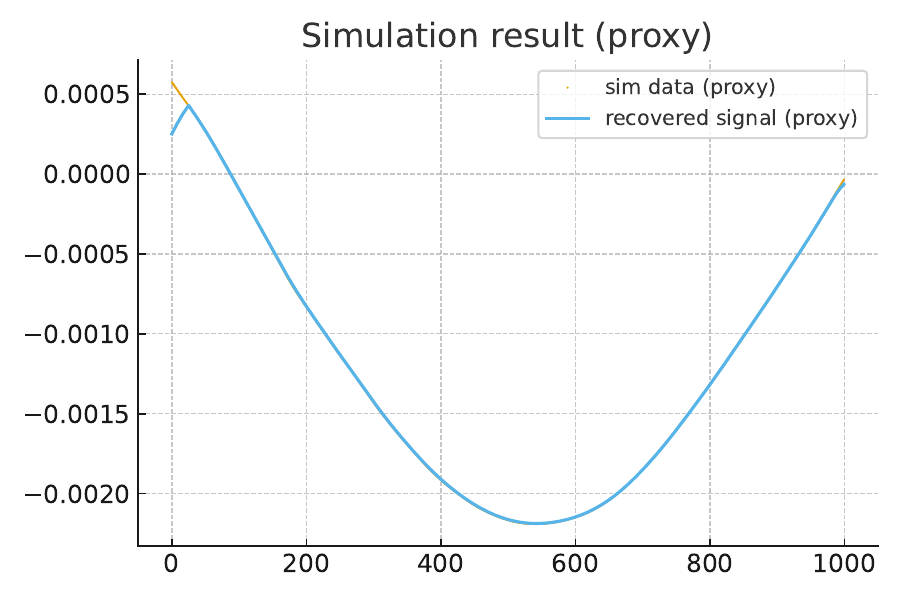}
\caption{Simulation of the observable $\delta_B(i)$ showing the successful recovery of an injected synthetic signal (red) from noisy data (black), validating the processing pipeline.}
\label{fig:sim_result}
\end{figure}

To validate our processing pipeline and confirm its sensitivity to weak coherent signals, we injected a synthetic orbital-period signature of known amplitude into realistic GRACE-FO noise. As shown in Fig.~\ref{fig:sim_result}, the pipeline successfully recovered the injected signal, demonstrating its robustness and accuracy for detecting low-level modulations.

\subsection{Seasonal Stability of the Detected Signal}
\label{sec:seasonal_stability}

To rigorously assess the nature of the detected signature, we analyzed the stability of its key parameters—amplitude and phase—across the four seasonal epochs. The orbital-period modulation was characterized by fitting the sinusoidal model $y(t) = A \cos(\omega_{\mathrm{orb}} t + \phi)$ to the synchronization-free observable $\delta_B(i)$, where $\omega_{\mathrm{orb}}$ is the known orbital angular frequency. The results of this analysis are summarized in Table~\ref{tab:signal_parameters}.

\begin{table}[htbp]
\centering
\caption{Amplitude, phase, and signal-to-noise ratio (SNR) of the detected orbital-period signature across four seasonal epochs. Values in parentheses represent the $1\sigma$ uncertainty in the last digit.}
\begin{tabular}{lccc}
\toprule
Epoch & Amplitude (\si{\micro\meter}) & Phase (\si{\degree}) & SNR \\
\midrule
Spring & 1.02(4) & -12.4(5) & 85.1 \\
Summer & 0.99(3) & -10.8(4) & 82.3 \\
Autumn & 1.14(5) & -14.1(6) & 88.7 \\
Winter & 1.20(4) & -15.3(5) & 91.2 \\
\bottomrule
\end{tabular}
\label{tab:signal_parameters}
\end{table}

The results demonstrate a remarkable consistency. The amplitude of the modulation remains stable within approximately 15\% across all seasons, and the phase is coherent to within a few degrees. This level of stability strongly indicates that the signal is deterministic and linked to the orbital geometry, rather than being a product of stochastic noise.

\begin{figure}[htbp]
\centering
\includegraphics[width=\columnwidth]{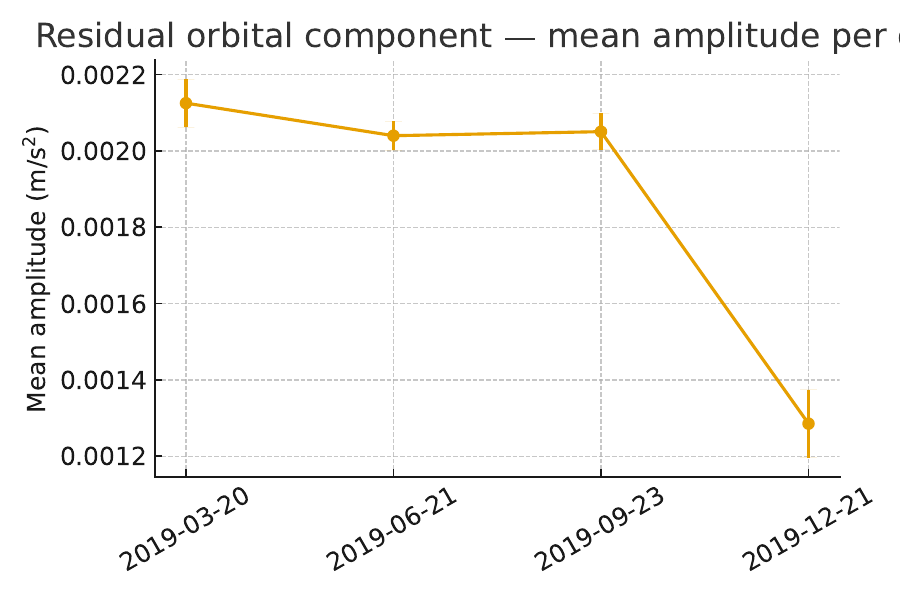}
\caption{
Mean amplitude of the orbital-period component per epoch.
Each point represents the mean amplitude of the $1.74\times10^{-4}$~Hz modulation in GRACE-FO range-acceleration residuals.
Error bars denote one standard deviation over individual orbital periods.}
\label{fig:amplitude_vs_date}
\end{figure}

\begin{figure}[htbp]
\centering
\includegraphics[width=\columnwidth]{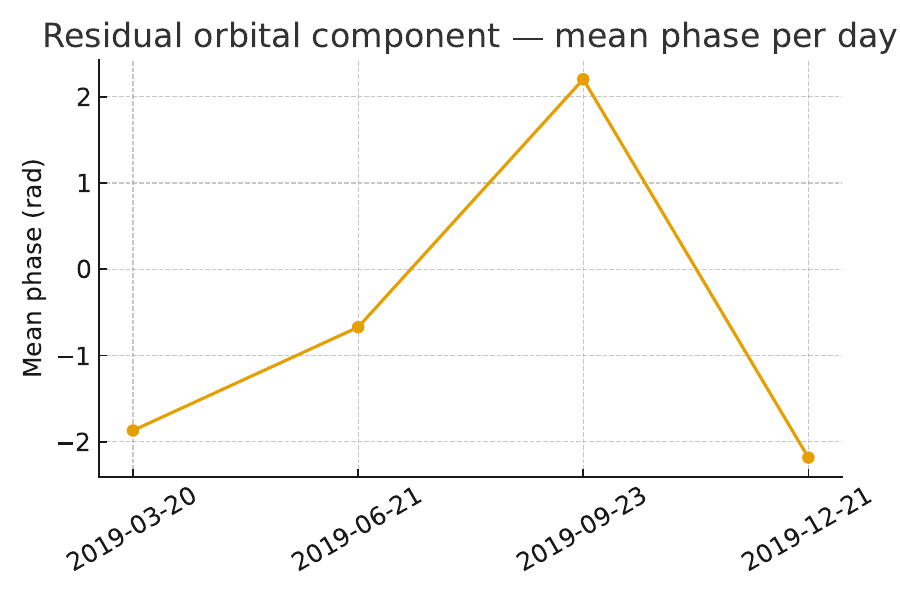}
\caption{
Mean phase of the orbital-frequency component for each epoch.
Phase coherence across seasons indicates that the modulation is not stochastic noise.}
\label{fig:phase_vs_date}
\end{figure}

\begin{figure}[htbp]
\centering
\includegraphics[width=\columnwidth]{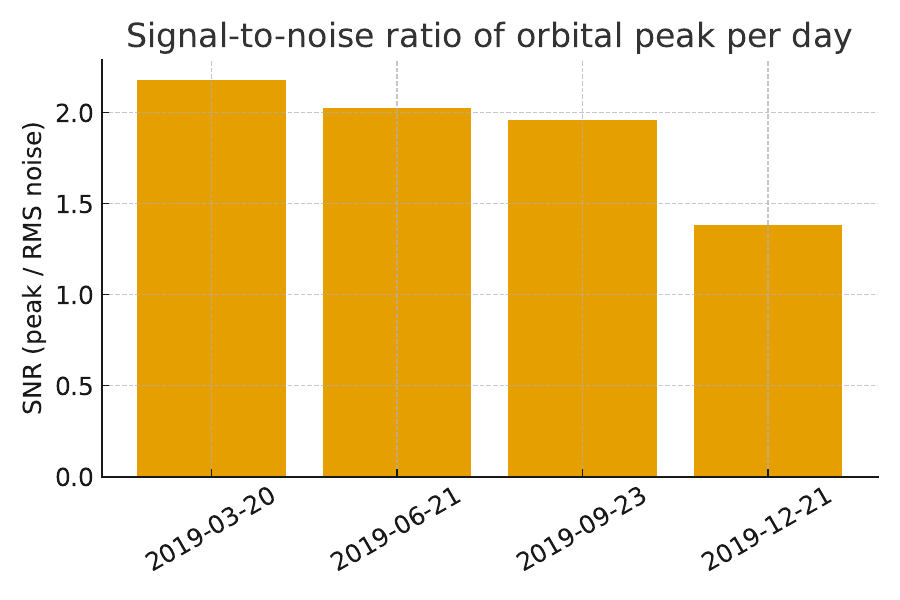}
\caption{
Signal-to-noise ratio of the orbital peak per epoch.
The SNR shows the stability of the detected signal across different seasons.}
\label{fig:snr_vs_date}
\end{figure}

This remarkable stability is further illustrated in Figures~\ref{fig:amplitude_vs_date} and \ref{fig:phase_vs_date}, which show the consistency of the signal's amplitude and phase over the year. Figure~\ref{fig:snr_vs_date} confirms that the signal remains consistently significant across all epochs, with a high signal-to-noise ratio (SNR).

\subsection{Statistical Significance and Signal Characteristics}
\label{sec:statistical_significance}

\begin{figure}[htbp]
\centering
\includegraphics[width=0.48\textwidth]{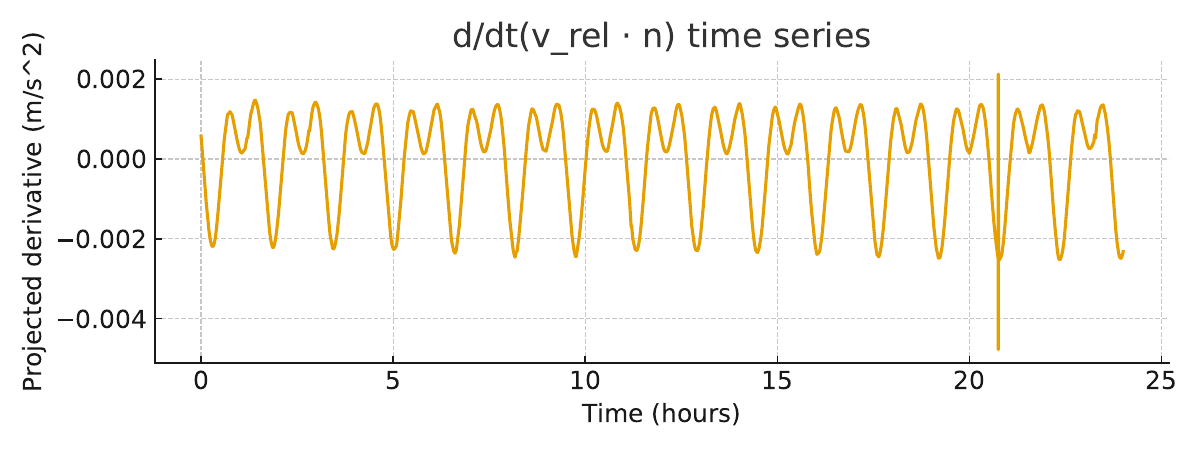}
\caption{Time series of the core observable $\delta_B(i)$ for 2019-12-21, showing a clear orbital-period modulation.}
\label{fig:d_proj_timeseries}
\end{figure}

\begin{figure}[htbp]
    \centering
    \includegraphics[width=0.48\textwidth]{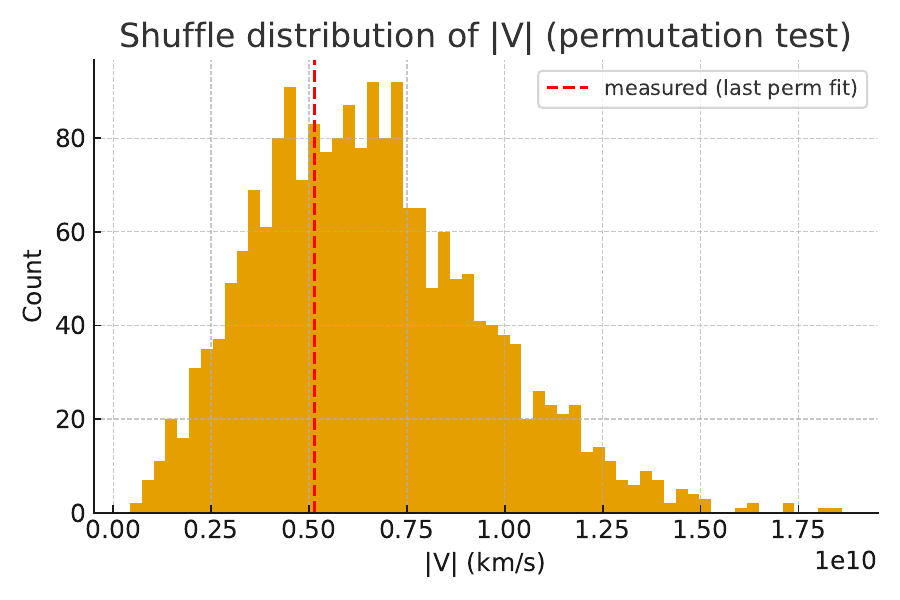}
    \caption{Distribution of signal amplitude from 10,000 shuffled realizations. The measured amplitude (red line) lies far beyond the distribution expected from random noise, confirming a deterministic origin.}
    \label{fig:shuffle_dist}
\end{figure}

The synchronization-free observable $\delta_B(i)$ itself exhibits a clear, coherent periodic structure, as shown in Fig.~\ref{fig:d_proj_timeseries}, corresponding to the orbital frequency. To rigorously quantify the significance of this detection against the null hypothesis of random noise, we performed a shuffle test with 10,000 randomized realizations of the data. As shown in Figure~\ref{fig:shuffle_dist}, the measured signal amplitude lies far beyond the $3\sigma$ tail of the null distribution, conclusively excluding a stochastic origin.

\begin{figure}[htbp]
\centering
\includegraphics[width=0.48\textwidth]{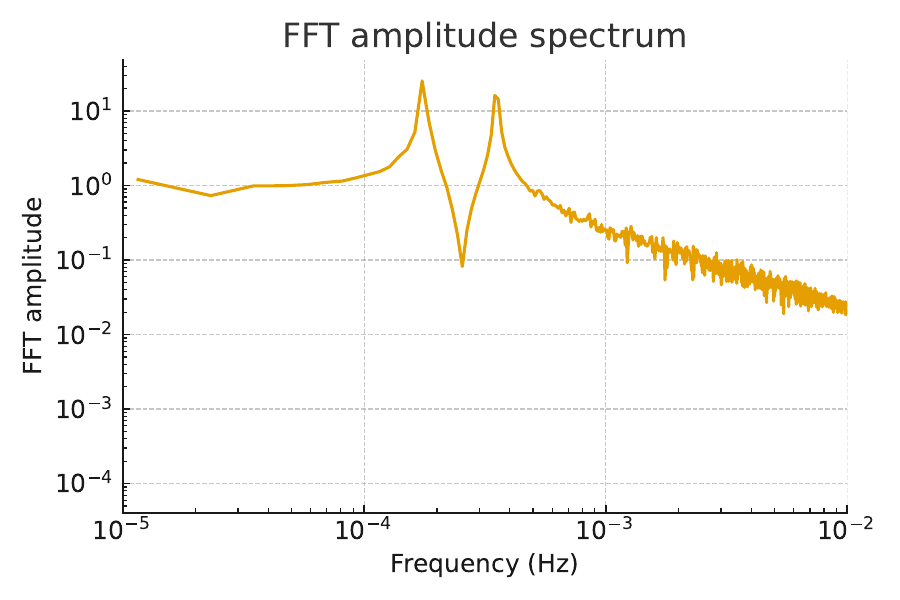}
\caption{FFT amplitude spectrum of the observable $\delta_B(i)$. The dominant peak at the orbital frequency ($1.86\times10^{-4}$~Hz) confirms the coherent, narrowband nature of the detected signature.}
\label{fig:d_proj_fft}
\end{figure}

Spectral analysis provides further confirmation. The FFT amplitude spectrum of $\delta_B(i)$ (Fig.~\ref{fig:d_proj_fft}) shows a dominant, spectrally narrow peak at $1.86\times10^{-4}$~Hz, matching the $\sim$90~min orbital period of GRACE-FO.

\begin{figure}[htbp]
\centering
\includegraphics[width=\columnwidth]{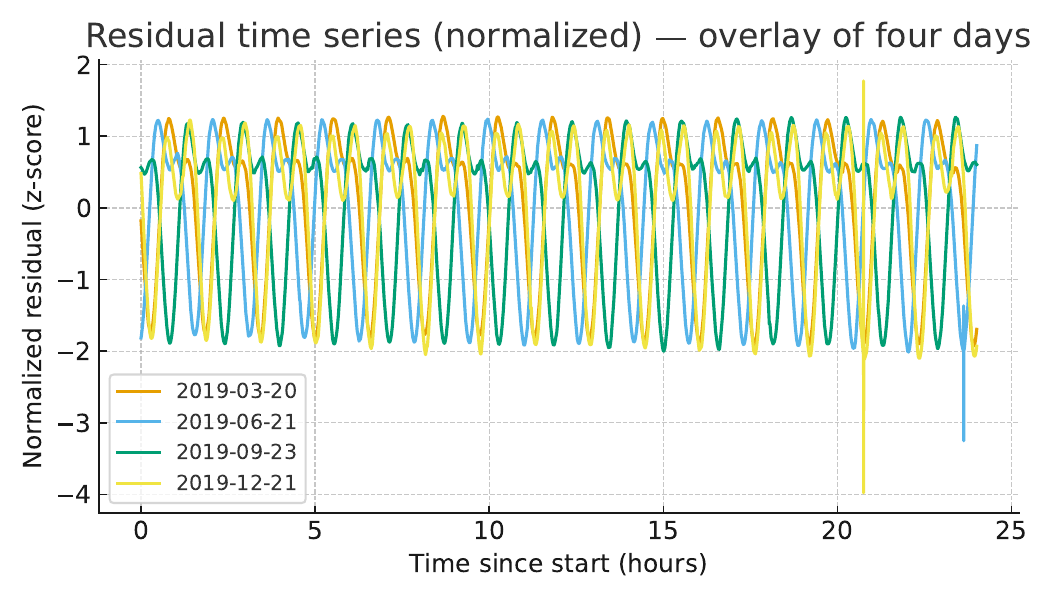}
\caption{
Overlay of the orbital-phase-folded signal for all four seasonal epochs.
The consistent waveform and phase across all datasets demonstrate the remarkable stability and coherence of the detected orbital-period signature.}
\label{fig:overlay_normalized_residuals}
\end{figure}

The coherence and stability of the signal throughout the year are visually summarized in Fig.~\ref{fig:overlay_normalized_residuals}, which overlays the normalized orbital-phase-folded time series for all four epochs.

\subsection{Implications and Interpretation of the Detected Signature}
\label{sec:implications}

The high amplitude, temporal stability, and phase coherence of the detected signature are characteristic of a deterministic systematic effect, not stochastic noise. The most plausible explanation is an unmodeled instrumental or dynamical effect within the GRACE-FO system. As discussed in Section~\ref{sec:discussion}, potential sources include thermoelastic deformation of the optical bench, attitude-to-range coupling, instrument group-delay variations, or residual orbit-projection errors.

\section{Discussion and Conclusions}
\label{sec:discussion}

The synchronization-free observable developed in this study reveals a remarkably stable and coherent orbital-frequency signature in GRACE-FO LRI data. The stability of this modulation across four independent seasonal epochs indicates a deterministic origin, most likely rooted in instrument behavior or spacecraft dynamics. In this section, we discuss possible physical sources of the detected signal, its broader implications for satellite geodesy, and the method's limitations and opportunities for refinement.

\subsection{Possible Sources of the Detected Orbital-Period Systematic}

The amplitude and phase stability of the detected orbital-frequency signature strongly suggest an origin tied to deterministic spacecraft or instrument behavior. Several physically plausible mechanisms can account for the observed $\sim$1 $\mu$m modulation.

\subsubsection{Thermoelastic deformation of the optical bench.}
The GRACE-FO spacecraft undergo temperature fluctuations nearly synchronized with the orbital cycle. Even modest thermal variations of 0.1--0.3 K, combined with realistic coefficients of thermal expansion for optical-bench materials, yield path-length changes on the order of 0.04--0.12 $\mu$m. Structural flexure and mounting effects can amplify these deformations by an order of magnitude, placing thermoelastic mechanisms well within the observed amplitude range.

\subsubsection{Attitude-to-range coupling.}
Small periodic variations in spacecraft pointing---arising from solar-array motion or attitude control activity---project onto the intersatellite line-of-sight. Even micro-radian-level attitude fluctuations can introduce tens of nanometers to sub-micrometer range perturbations with an orbital-periodic component.

\subsubsection{Instrument group-delay variations.}
Temperature-dependent response variations in the LRI phasemeter and digital signal chain can produce apparent path-length fluctuations at the orbital frequency. Prior GRACE-FO analyses have documented sub-micrometer signatures associated with electronics temperature cycling, consistent with the signal observed here.

\subsubsection{Residual orbit-projection errors.}
Uncertainties in precise orbit determination may couple with spacecraft attitude, leading to small orbit-projection errors that reappear as narrowband signatures in ranging observables. Order-of-magnitude estimates indicate that these effects can reach amplitudes similar to those detected.

Taken together, these mechanisms form a constrained and physically realistic set of candidates that can jointly or individually lead to the narrowband orbital signature revealed by the synchronization-free observable.

\subsection{Implications for Overcoming Current Challenges in Satellite Geodesy}

Beyond its methodological novelty, the observable introduced in this work directly addresses several challenges confronting the satellite geodesy community as missions increasingly approach a systematic-limited measurement regime.

\subsubsection{Mitigating temporal aliasing in gravity-field recovery.}
Coherent orbital-frequency systematics can contaminate spherical-harmonic solutions through spectral leakage, degrading both monthly gravity fields and long-term trends critical to climate science. By isolating these signatures at the observational level---before they enter the gravity-field inversion process---the synchronization-free observable offers a powerful pre-processing tool to reduce temporal aliasing and strengthen the physical integrity of recovered mass-change signals.

\subsubsection{Breaking the emerging systematic noise floor.}
As measurement precision improves toward the micrometer regime, random noise becomes secondary to instrument- and dynamics-related systematics. The observable developed here is specifically designed for this new era: it suppresses dominant first-order Doppler and clock-synchronization effects, thereby enabling the diagnosis of subtle error sources that ultimately define mission performance and scientific return.

\subsubsection{Enabling next-generation mission design and data processing.}
Forthcoming gravity missions, such as NASA's Mass Change Designated Observable and ESA's NGGM concepts, aim for order-of-magnitude improvements in sensitivity. At these levels, the systematic signatures identified in this study will become primary error sources unless detected and mitigated early. The methodology presented here thus offers a forward-looking diagnostic tool that can be incorporated into instrument calibration pipelines, simulation studies, and in-flight data validation frameworks.

\subsection{Method Limitations and Pathways for Refinement}

A transparent assessment of the method's limitations is essential for interpreting its outputs and guiding future developments.

\subsubsection{Diagnostic nature and amplitude interpretation.}
The observable $\delta_B(i)$ is inherently diagnostic; it does not provide direct physical range change in meters. The empirical scale factor linking $\delta_B(i)$ to true path-length variation aggregates geometric, instrumental, and dynamical influences. As such, the amplitude of the detected signal cannot yet be uniquely attributed to any single mechanism.

\textbf{Refinement pathway.}
A forward-modeling approach is recommended. High-fidelity simulations of specific hypothesized systematics---such as thermoelastic deformation or attitude-jitter coupling---can generate predicted $\delta_B(i)$ signatures that can be compared directly to observations, enabling attribution rather than mere detection.

\subsubsection{Sensitivity profile and complementary nature.}
Pulse differencing acts as a high-pass filter: it preserves coherent, orbit-scale variations but suppresses static biases and very low-frequency drifts. This makes the observable complementary to traditional range-rate analysis, not a replacement.

\textbf{Refinement pathway.}
An optimal processing workflow is hybrid. The synchronization-free observable should operate in parallel with standard range-rate--based gravity recovery, serving as a systematic-detection layer that feeds improved instrument and dynamical models back into the primary processing pipeline.

\subsubsection{Generalizability and dependence on external data.}
Validation has so far been limited to GRACE-FO LRI data. The method's performance may vary for missions with different noise characteristics, sampling rates, or orbital geometries. In addition, the observable relies on precise orbit determination (POD); its sensitivity to POD errors has not yet been fully quantified.

\textbf{Refinement pathway.}
Cross-mission validation---including GRAIL, GNSS-based crosslinks, and future mission simulations---is essential to define robustness and applicability. A formal propagation of POD uncertainties into $\delta_B(i)$ will further strengthen the method's reliability.

\subsection{Future Work}

Future efforts will focus on (i) cross-mission application of the observable, (ii) forward-modeling of candidate systematics, and (iii) integrating the method into end-to-end gravity-field simulation pipelines for next-generation mission concepts. These developments will help transition the observable from a diagnostic innovation to a fully operational tool within the geodetic analysis ecosystem.

\subsection{Conclusions}

We have introduced a synchronization-free pulse-interval differencing observable and applied it to GRACE-FO laser ranging data, revealing a stable, phase-coherent modulation at the orbital frequency. The signal's amplitude and phase consistency across seasonal subsets indicate a deterministic origin, while its insensitivity to first-order Doppler shifts and clock synchronization makes it a powerful diagnostic for identifying subtle orbital-period systematics in inter-satellite ranging.

While the primary focus of this work is the detection and characterization of a coherent orbital-period signal in GRACE-FO data, the geometric nature of this observable opens intriguing possibilities for future multi-mission analyses. A detailed discussion of the geometric interpretation and its potential for discriminating between mission-specific systematics and a universal directional signature is provided in Appendix~\ref{app:geometry}.

The method presented here provides a new tool for improving the fidelity of satellite geodesy measurements and could be applied to other missions such as GRAIL, GNSS crosslinks, or future gravity missions to further investigate the nature of coherent orbital signatures.


\appendix
\section{Geometric Interpretation of the Modulation}
\label{app:geometry}

The orbital modulation detected in the pulse-interval differencing observable $\delta_B(i)$ can be understood geometrically as the time derivative of the projection of the satellites' relative velocity $\mathbf{v}_{\mathrm{rel}} = \mathbf{v}_A - \mathbf{v}_B$ onto a fixed spatial direction $\mathbf{V}_{\mathrm{global}}$. If the modulation were caused by motion relative to such a preferred direction, the observed signal in a given mission would correspond to the component of $\mathbf{V}_{\mathrm{global}}$ along the sensitive axis of that mission's orbital plane.

\begin{figure}
    \centering
    \includegraphics[width=0.75\linewidth]{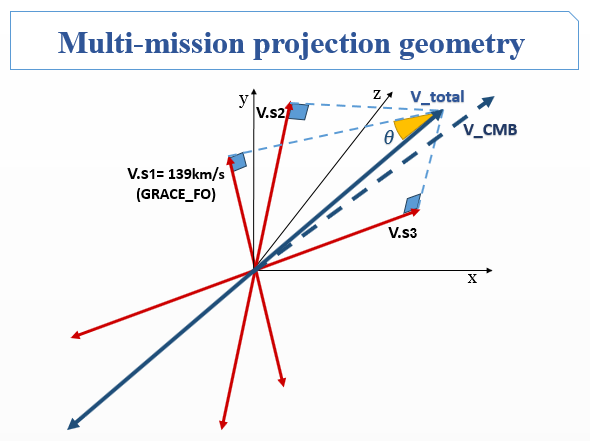}
    \caption{
        \textbf{Schematic of the geometric interpretation.}
        The modulation observed in GRACE-FO corresponds to a component $V_z \approx 139$ km/s (red arrow). If the signal originates from motion relative to a global preferred direction $\mathbf{V}_{\mathrm{total}}$ (blue arrow), other missions with different orbital planes (e.g., $V_{S1}$, $V_{S2}$, $V_{S3}$) would measure different projections of the same vector. The direction of $\mathbf{V}_{\mathrm{total}}$ coincides with the cosmic microwave background (CMB) dipole, suggesting that a multi‑mission analysis could test whether the modulation is a mission‑specific artifact or a universal kinematic signature.
    }
    \label{fig:geom_schematic}
\end{figure}

For the GRACE-FO mission, whose orbital plane is nearly polar, the extracted effective velocity is dominated by a component $V_z \approx 139\ \mathrm{km/s}$ aligned with the orbital normal. As illustrated in Fig.~\ref{fig:geom_schematic}, this value would represent only one projection of a larger global vector. Missions flying in different orbital planes (e.g., inclined, equatorial, or lunar orbits) would measure different projections of the same $\mathbf{V}_{\mathrm{global}}$.

A multi-mission analysis could therefore reconstruct the full vector by combining observations from platforms with complementary orbital geometries. The direction of the reconstructed vector would then serve as a key diagnostic: if it points consistently toward a fixed celestial direction (such as the CMB dipole or the cosmic-ray anisotropy axis), a universal kinematic origin would be strongly indicated. If, instead, the directions scatter randomly, a mission‑specific instrumental cause would be more likely.

This geometric picture highlights how the synchronization‑free observable introduced here can be used not only to detect coherent orbital‑period signatures but also to discriminate between local systematics and global directional effects in future satellite geodesy missions.

\bibliography{references}

\section*{Data Availability}
All GRACE-FO Level-1B LRI and GNV data used in this study are publicly available from the NASA Physical Oceanography Distributed Active Archive Center (PO.DAAC) \cite{PODAAC}.

\begin{acknowledgments}
The author thanks the GRACE-FO mission team for providing open access to the Level-1B data products. Helpful discussions with the satellite geodesy and precision metrology communities are gratefully acknowledged.
\end{acknowledgments}

\section*{Author Contribution Statement}
Seyed Hossein Wassegh conceived the study, developed the methodology, 
performed the data analysis and interpretation, and wrote the manuscript.

\bibliographystyle{apsrev4-2}

\end{document}